\documentclass[
]{ceurart}

\usepackage{listings}
\usepackage{natbib}

\usepackage{mwe}
\usepackage{subfig}
\begin{document}

\copyrightyear{2024}
\copyrightclause{Copyright for this paper by its authors.
  Use permitted under Creative Commons License Attribution 4.0
  International (CC BY 4.0).}

\conference{AI for Access to Justice Workshop co-located with Jurix 2024, Brno, Czech Republic}

\title{LLMs \& Legal Aid: Understanding Legal Needs Exhibited Through User Queries}

\author[1]{Michal Kuk}[%
email=michal.kuk@frankbold.org,
]
\cormark[1]
\fnmark[1]
\address[1]{Frank Bold Society, z.s., Brno, Czech Republic}

\author[2]{Jakub Harasta}[%
orcid=0000-0002-5722-0325,
email=harasta@muni.cz,
]
\fnmark[1]
\address[2]{Institute of Law and Technology, Faculty of Law, Masaryk University, Brno, Czech Republic}

\cortext[1]{Corresponding author.}
\fntext[1]{These authors contributed equally.}

\begin{abstract}
   The paper presents a preliminary analysis of an experiment conducted by Frank Bold, a Czech expert group, to explore user interactions with GPT-4 for addressing legal queries. Between May 3, 2023, and July 25, 2023, 1,252 users submitted 3,847 queries. Unlike studies that primarily focus on the accuracy, factuality, or hallucination tendencies of large language models (LLMs), our analysis focuses on the user query dimension of the interaction. Using GPT-4o for zero-shot classification, we categorized queries on (1) whether users provided factual information about their issue (29.95\%) or not (70.05\%), (2) whether they sought legal information (64.93\%) or advice on the course of action (35.07\%), and (3) whether they imposed requirements to shape or control the model's answer (28.57\%) or not (71.43\%). We provide both quantitative and qualitative insight into user needs and contribute to a better understanding of user engagement with LLMs. 
\end{abstract}


\maketitle

\section{Introduction}
\label{introduction}

This paper reports on the experiment running between May 3, 2023, and July 25, 2023, which utilized GPT-4 to answer legal questions submitted to Frank Bold Legal Counseling Center in the Czech Republic. The paper does not address the accuracy and factuality of the model's answers. Instead, we focus on analyzing queries, allowing a unique insight into legal needs demonstrated by users intuitively interacting with GPT-4.

ChatGPT was launched on November 30, 2022, drawing significant attention to the natural language capabilities of large language models (LLMs). Its accessibility immediately became a major disruptive force, raising questions about its potential impact on legal services and access to justice. Various organizations worldwide have become interested in leveraging LLMs to support, scale, or restructure their operations.

Frank Bold (FB), a Czech expert group offering for-profit and non-profit services, has experimented with ChatGPT since early 2023. FB ran a public experiment between May 3, 2023, and July 25, 2023, mediating access to ChatGPT to the general public seeking legal aid. Throughout the experiment, 1,252 users submitted 3,847 queries, to which GPT-4 responded. Various studies focused on capabilities displayed by models \citep{Tan2023, Deroy2023, Deroy2024, Ramprasad2024, Westermann2023Mediator, Westermann2023Layperson, Goodson2023}, and prevalence of hallucinations \citep{Wang2023, Magesh2024}. However, the issue of AI/GAI/LLM's ability to perform human tasks is largely detached from user expectations and perceptions related to interactions with LLMs. We aim to support surveys by \citet{Hagan2024} and \citet{Cheong2024} and provide a better understanding of user legal needs by analyzing larger queries arising from their intuitive use of GPT-4.

We report on statistics of the user queries and offer insight into the legal needs the users intuitively manifested throughout the experiment. We outline the experiment (Section \ref{experiment}), offer quantitative statistics of the user queries (Section \ref{statistics}), and qualitative analysis of the observed trends (Section \ref{topics}). The discussion (Section \ref{discussion}) addresses our findings in the context of existing literature focused on user needs and related risks. As the work is preliminary and descriptive, we outline the future work required to explore the issue in more detail (Section \ref{future}).

\section{Experimental set-up}
\label{experiment}

In 2023, Frank Bold conducted an experiment to explore and evaluate the potential applications of the newly accessible GPT-4 model in the context of legal aid. Frank Bold is an expert group established in 2013 as a collective of entities offering both for-profit and non-profit services in law and other areas. The services include a commercial law firm and a non-profit online Legal Counseling Center (\textit{Právní poradna} in Czech). The latter provides the Czech public with legal information and assistance in areas of public interest, including environmental law, whistleblowing and corruption-related issues, civic rights, municipal laws, and civic engagement issues. Additionally, the counselling centre offers legal technology tools such as interactive interviews on legal matters and document generators.

The experiment utilizing GPT-4 was initiated on May 3, 2023, and was accessible via the now (December 2024) defunct online platform at \url{www.ai.frankbold.org}. The effort was presented as an exploration of the experimental application of artificial intelligence to address legal questions. Initially, given the internal funds available, the limit was set at providing answers to 3,000 user queries. On June 19, 2023, the limit was increased to 4,000 user queries. The limit was reached on June 10, 2023, when the experiment was concluded. The experiment was made public through FB's internal mailing lists. As the first effort of its kind, information about the project was disseminated through several prominent online media outlets.

Users interested in submitting their queries needed to create a user account and provide a valid e-mail address and a full name. Additionally, users could, as non-required information, provide additional information such as their profession, the organization to which they were affiliated, and their phone number. Once the registration and login processes were complete, users could submit their queries via an interface comprising a single input form. The tool operated on a single question-single answer basis. It did not facilitate chat-like interaction, which is now well-known to users of ChatGPT, and did not allow follow-up questions. Users were limited to 10 questions per day. Users could select whether to wait for the answer (being informed that it might take up to 3 minutes to provide it) or prefer the answer sent to their e-mail (the e-mail option was added later in the experiment on May 25). The delay was caused by the set-up presenting the user with the final answer created by the LLM instead of having it in a well-known streaming text format. 24\% of the users preferred receiving answers via e-mail. Subsequently, the answer was enriched by links to relevant articles from FB's Legal Counseling Center and the FB Law Firm's blog, which could be accessed to gather further, more detailed information. The final element was a voluntary option to rate the provided answer and provide textual feedback.

The LLM-generated answer employed retrieval-augmented generation (RAG) comprising the following steps:
\begin{enumerate}
    \item Retrieval of user query.
    \item Identification of the relevant context through a similarity search between the user's query and FB's guidelines, blog posts, and articles. Later in the experiment, on May 25, selected legal acts were added as a context source.
    \item Selection of the relevant context based on similarity search.
    \item Combination of the relevant context with user query to form a single prompt.
    \item Retrieval of answer generated by GPT-4.
    \item Selection of links to guidelines, blog posts, and articles based on similarity search of the answer.
    \item Presentation of the answer to the user on the website or via e-mail.
\end{enumerate}

To provide users with guidance and to manage expectations, a comprehensive set of instructions and a detailed disclaimer were displayed to the user. Before submitting a query, the user was informed that the tool performs optimally in the subject areas where FB's Legal Counseling Centre has developed a particular expertise. These areas were outlined. Additionally, the users were made aware of the tool's inability to draft documents or look up online information. Once users received the answer, they were reminded that the tool is experimental, and the answers provided cannot be trusted without verification, ideally provided by an experienced attorney.

Throughout the experiment, some of the internal variables changed. For instance, different versions of the GPT-4 model were used. Additionally, details of the RAG process were adjusted, leading to the use of different approaches to select contexts or benchmarks for similarity searches. The last parameter adjusted during the experiment was the system message, which set instructions for the user's query. These adjustments were motivated by FB's need to test different strategies and familiarize themselves with the novel technology.

Ultimately, the goal of the experiment was the exploration of LLM's capabilities to boost both for-profit and non-profit services provided by FB. Frank Bold conducted the experiment independently, with the paper's first author (M.K.) being a principal investigator and developer. The data collection observed legal requirements and FB's ethical policy. Only after the data were collected did the second author (J.H.) participate in the data analysis. A strict policy was implemented to avoid the second author's access to personal data and potentially personally identifiable information.

\section{Numbers…}
\label{statistics}
\begin{figure}
    \centering
    \includegraphics[width=1\linewidth]{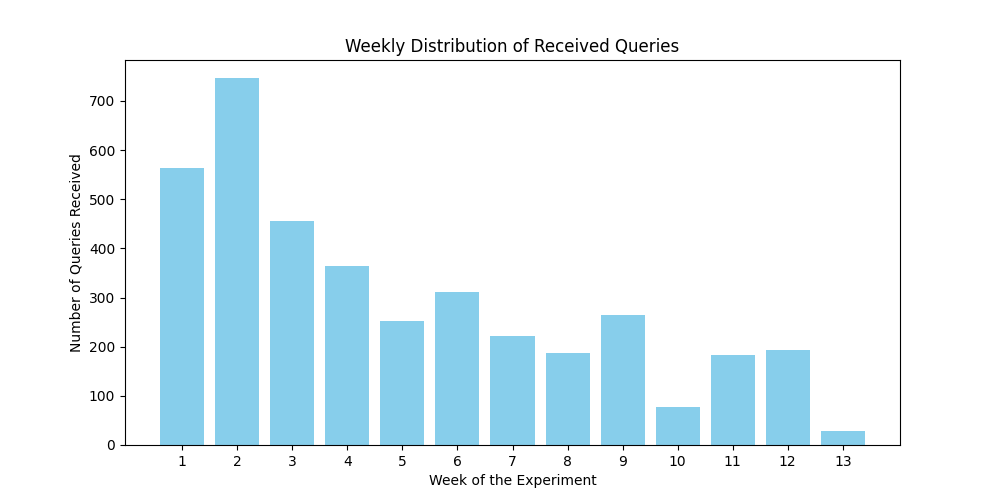}
    \caption{Weekly Distribution of Queries}
    \label{fig:weekly-distribution}
\end{figure}

Throughout the experiment, a total of 4,045 queries were submitted by 1,262 registered users. The total number of users who registered for the experiment was 1,543. However, 281 (18\%) of the registered users never submitted any query. 72\% of the queries were submitted in the first half of the 13 weeks of the experiment running. Weekly distribution is presented in Figure \ref{fig:weekly-distribution}. Before the analysis, we removed duplicities in queries and queries that were out-of-scope of the experiment (such as users asking for flight duration between New York and Prague, etc.). The preprocessing brought the final tally to 3,847 queries submitted by 1,252 users.

\begin{figure}
    \centering
    \includegraphics[width=1\linewidth]{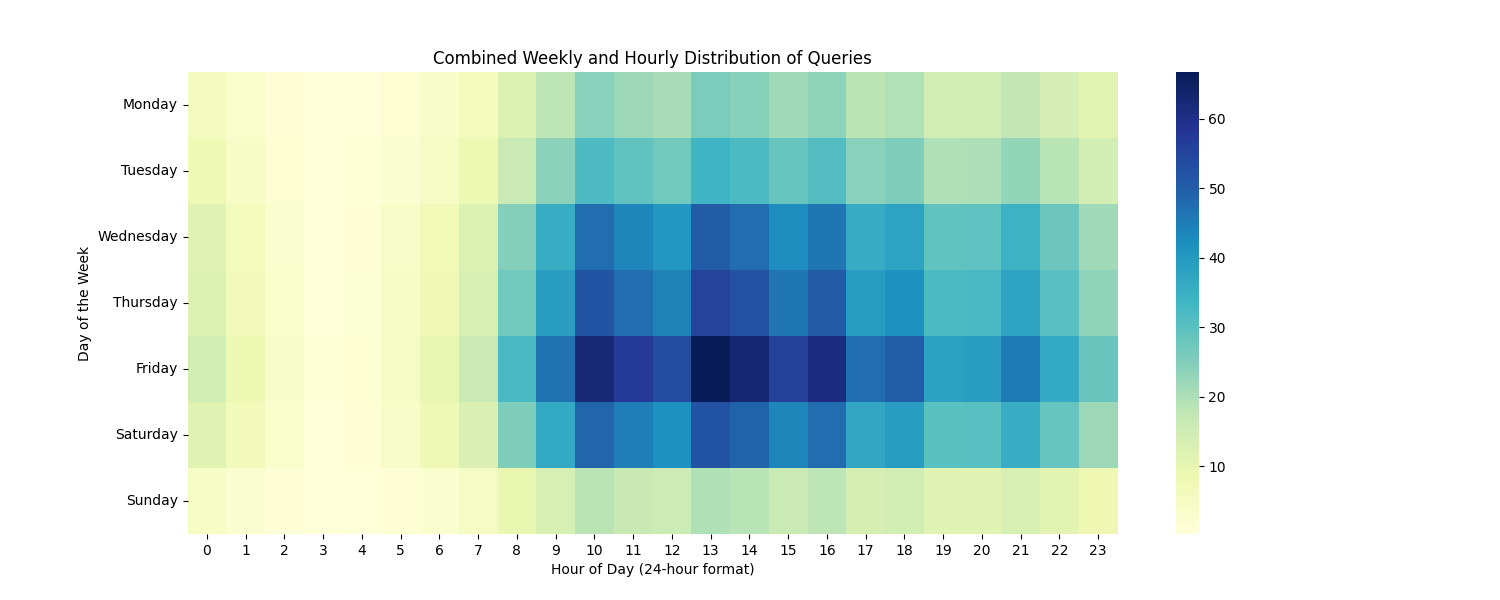}
    \caption{Distribution of Queries by Day of Week and Time}
    \label{fig:distribution-in-week}
\end{figure}

One of the benefits of online legal assistance tools, including those that employ AI, is their accessibility at any time. In this experiment, most queries were submitted on Friday, with 61\% of the queries submitted between 9:00 and 17:00 and 76\% submitted during workdays. Therefore, a substantial part of queries were submitted during standard work hours, but with a clear preference for days later in the week rather than its beginning, as shown in Figure \ref{fig:distribution-in-week}.

The length of the queries exhibited considerable variability, spanning from 5 to 8499 characters. The average length was nearly 224 characters, and the median was 117 characters (see Table \ref{tab:length}). Notably, there was no discernible correlation between the query length and the answer length (r = 0.059). The lack of correlation may be attributed to a system prompt that was present, which provided the LLM with information regarding the structure of the desired response. Additionally, no correlation was identified between the number of previous questions submitted by a user and the length of the question (r = 0.023).

\begin{table}
  \caption{Query Length Statistics}
  \label{tab:length}
  \begin{tabular}{ccl}
    \toprule
    \textbf{Metric} & \textbf{Value} \\
    \midrule
    Average & 223.85 \\
    Median & 117 \\
    Min & 5 \\
    Max & 8499 \\
    Over 2000 characters & 33 \\
    \bottomrule
  \end{tabular}
\end{table}

\begin{figure}
    \centering
    \includegraphics[width=1\linewidth]{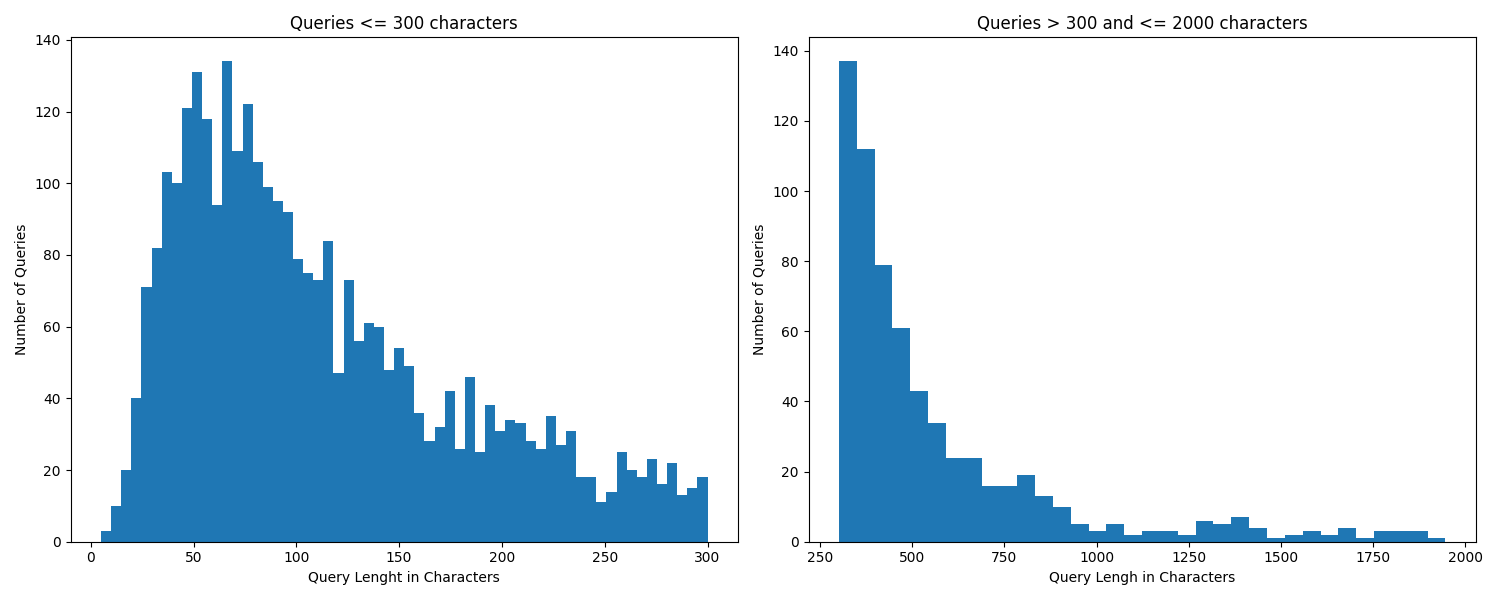}
    \caption{Query Length Distribution}
    \label{fig:query-length}
\end{figure}

The shortest queries were one-to-three-word search queries. They included terms such as "rent," "divorce," "acknowledgement of debt," and other queries that closely resembled standard online search queries. In contrast, the longest prompts were typically due to individuals including text from documents, histories of email communication, or other contextual information in their queries. Both instances were present in the queries but were not usual, as seen in Figure \ref{fig:query-length}.

Of 1,252 users who asked any relevant question 55\% submitted exactly one query. The remaining respondents submitted at least two responses, with 11\% of users submitting five or more queries. See Table \ref{tab:commands} overview. Such a distribution indicates either a desire for further information or recognition of the value of the initial response and a willingness to use the model again.

\begin{table}
  \caption{Number of Questions Submitted by Users}
  \label{tab:commands}
  \begin{tabular}{ccl}
    \toprule
     \textbf{Number of queries by user}& \textbf{Percentage of users}\\
    \midrule
    One query & 55\% \\
    Two to five queries & 34\% \\
    Five to ten queries & 6\% \\
    More than ten queries & 5\% \\
    \bottomrule
  \end{tabular}
\end{table}

\section{… and needs}
\label{topics}

It is safe to assume that there are human needs behind every submitted query. The users facing legal issues and having legal needs approached the interface with some level of expectation that their needs would be met. Our understanding of user interaction with the LLM is often shaped by expert-driven \citep{Cheong2024} or community-driven findings \citep{Hagan2024} based on workshops and interviews. To our knowledge, no analysis provides insight into user needs by observing an intuitively used set of queries submitted by users trying to obtain legal information or legal help.

One of the main concerns about using GPT-4 for legal help and self-help is the blurry line between obtaining legal information and seeking legal advice. Providing legal opinions is associated predominantly with human experts, who are, as members of the legal profession, accountable and liable for their advice.

\citet{Cheong2024} proposed a framework describing the issues related to potential risks and avenues to consider when developing responsible LLM policies for legal advice. One of the dimensions identified in their paper were considerations about the user queries dissipating into three interconnected parts: assessment of facts, identification of relevant laws, and the nature of desired answers. Their analysis serves as our starting point for classifying user queries submitted throughout the experiment.

We randomly selected 200 queries from the dataset of user-submitted queries and developed descriptive codes to help us understand their nature.

Through an iterative process, and based on Cheong et al., we developed the following categories \footnote{Queries and prompts were in the Czech language. The description of categories is translated into English for the purpose of the paper.}:

\begin{enumerate}
    \item \textbf{Facts in User Query}: Queries can describe facts about a specific situation the user encountered. Users tend to either explain the facts or submit a specific document. Alternatively, queries can contain no facts.
    \item \textbf{Information about the Law}: Queries can seek retrieval of specific legal information or identification and lookup of a specific act or case law. Alternatively, queries seek advice on further course of action and possible solutions.
    \item \textbf{User Grants Control}:  Queries can pose open-ended questions, which provide no guidance in structuring the answer and grant control over the answer to the model. Alternatively, queries can be formulated to impose requirements on the answer's structure or format.
\end{enumerate}

We pose that queries providing facts, seeking advice, and containing open-ended questions manifest intuitive users' expectations of obtaining personalized and actionable advice about the further course of action and treat the GPT-4 akin to a human expert. Alternatively, queries which do not provide facts, seek information, and limit the possible answer space manifest intuitive user expectations to obtain information about the law and treat the GPT-4 as a sophisticated search engine.

We employed zero-shot classification to analyze the prevalence of these codes in the dataset of user queries. We provided category descriptions as prompts and employed GPT-4o for classification. We did not evaluate the outcome of zero-shot classification.

The classification yielded 1,153 (29.95\%) queries as containing facts, and 2,694 (70.05\%) queries not containing any facts that users tried to present to GPT-4. Regarding information patterns, 2,498 (64.93\%) queries sought information, and 1,349 (35.07\%) queries sought advice on further course of action. Finally, 2,748 (71.43\%) queries contained open-ended questions. Therefore, in the case of  1,099 (28.57\%) queries, users tried to limit the decision space and maintain control over the answer granted by the LLM.

\begin{figure}

\begin{minipage}{.5\linewidth}
\centering
\subfloat[]{\label{main:a}\includegraphics[scale=.5]{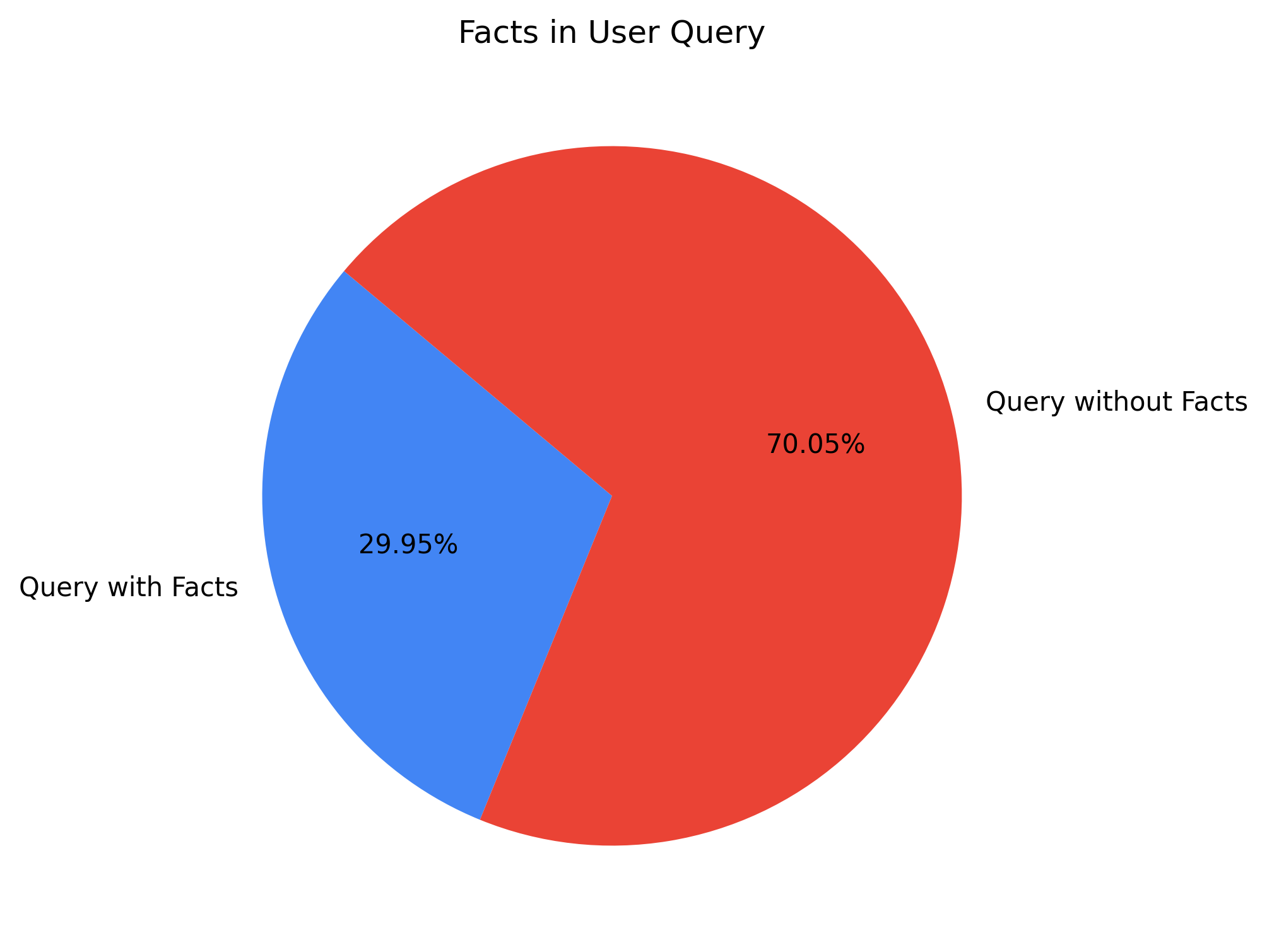}}
\end{minipage}%
\begin{minipage}{.5\linewidth}
\centering
\subfloat[]{\label{main:b}\includegraphics[scale=.5]{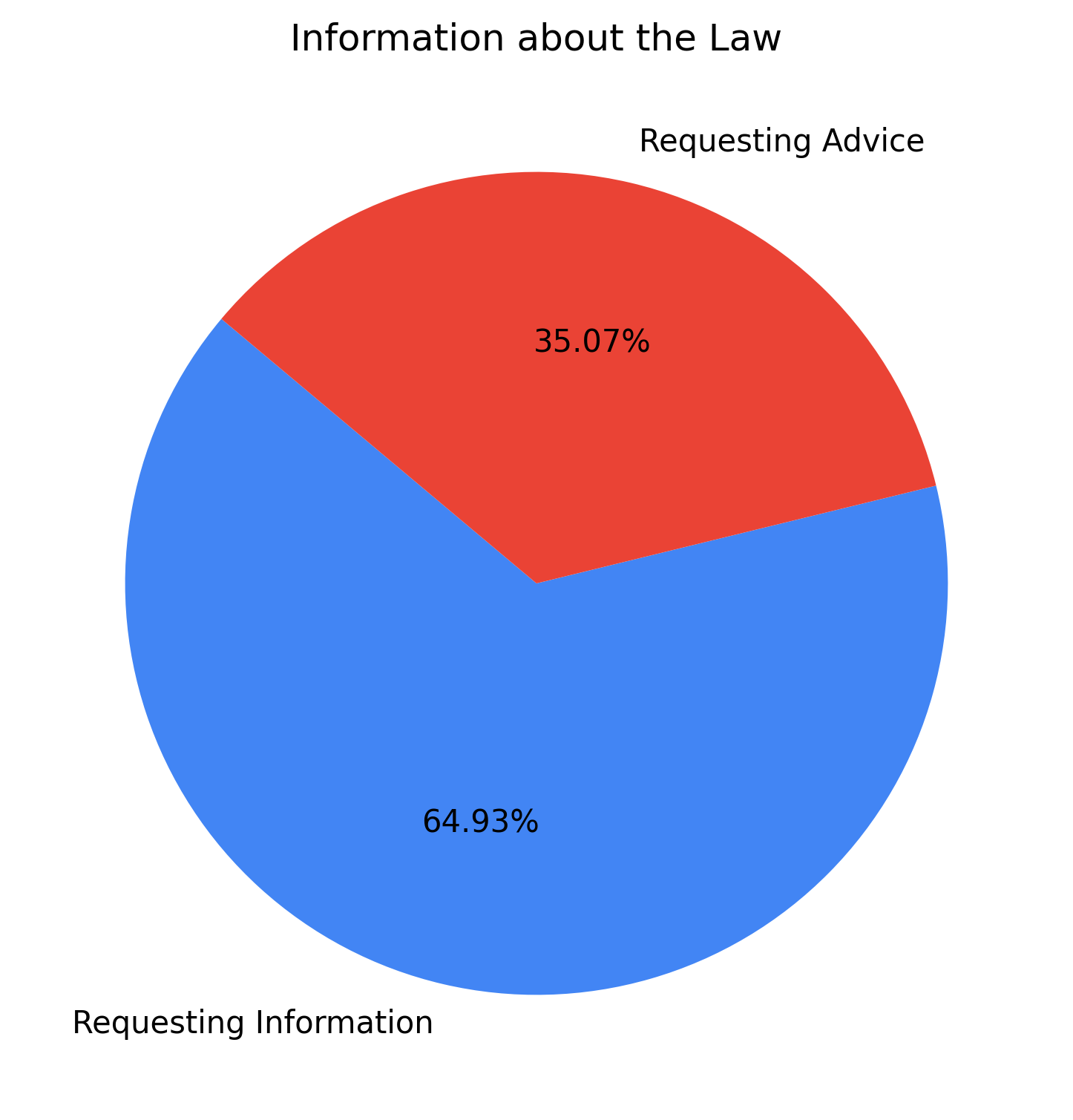}}
\end{minipage}\par\medskip
\centering
\subfloat[]{\label{main:c}\includegraphics[scale=.5]{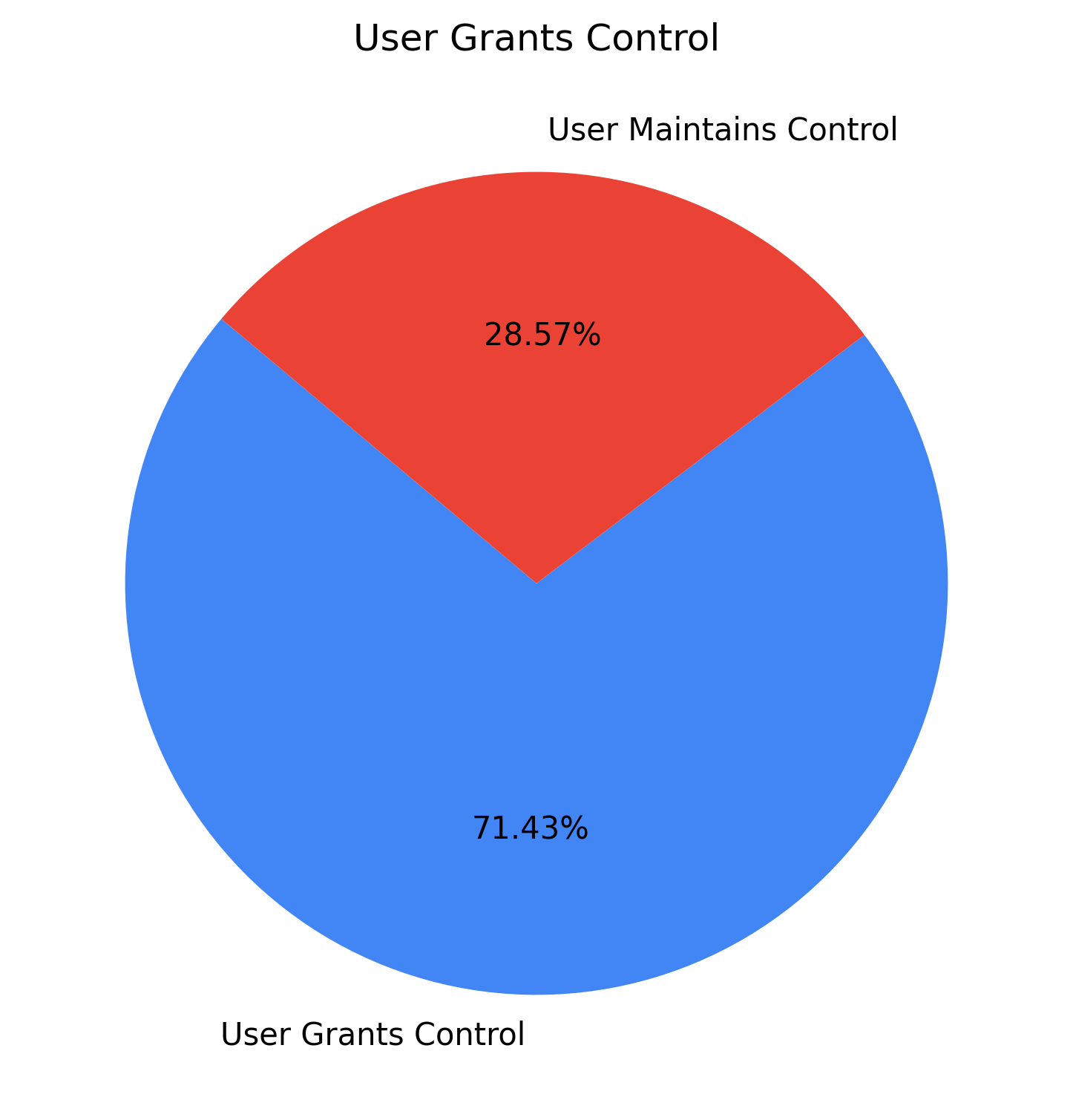}}

\caption{Distribution of Codes (Zero-Shot Classification using GPT-4o; no subsequent evaluation conducted)}
\label{fig:codes}
\end{figure}

We found that only 129 queries expected fully-pledged and personalized actionable advice. Meaning that the query, cumulatively, offered facts, did not seek retrieval of information about the law (but advice instead), and did not try to limit the answer provided by the model. On the other hand, we found that only 117 queries expected a look-up. This means the query did not offer facts to the system, sought retrieval of information about the law, and maintained control by imposing structural or other limitations on the expected answer.

\begin{figure}
    \centering
    \includegraphics[width=1\linewidth]{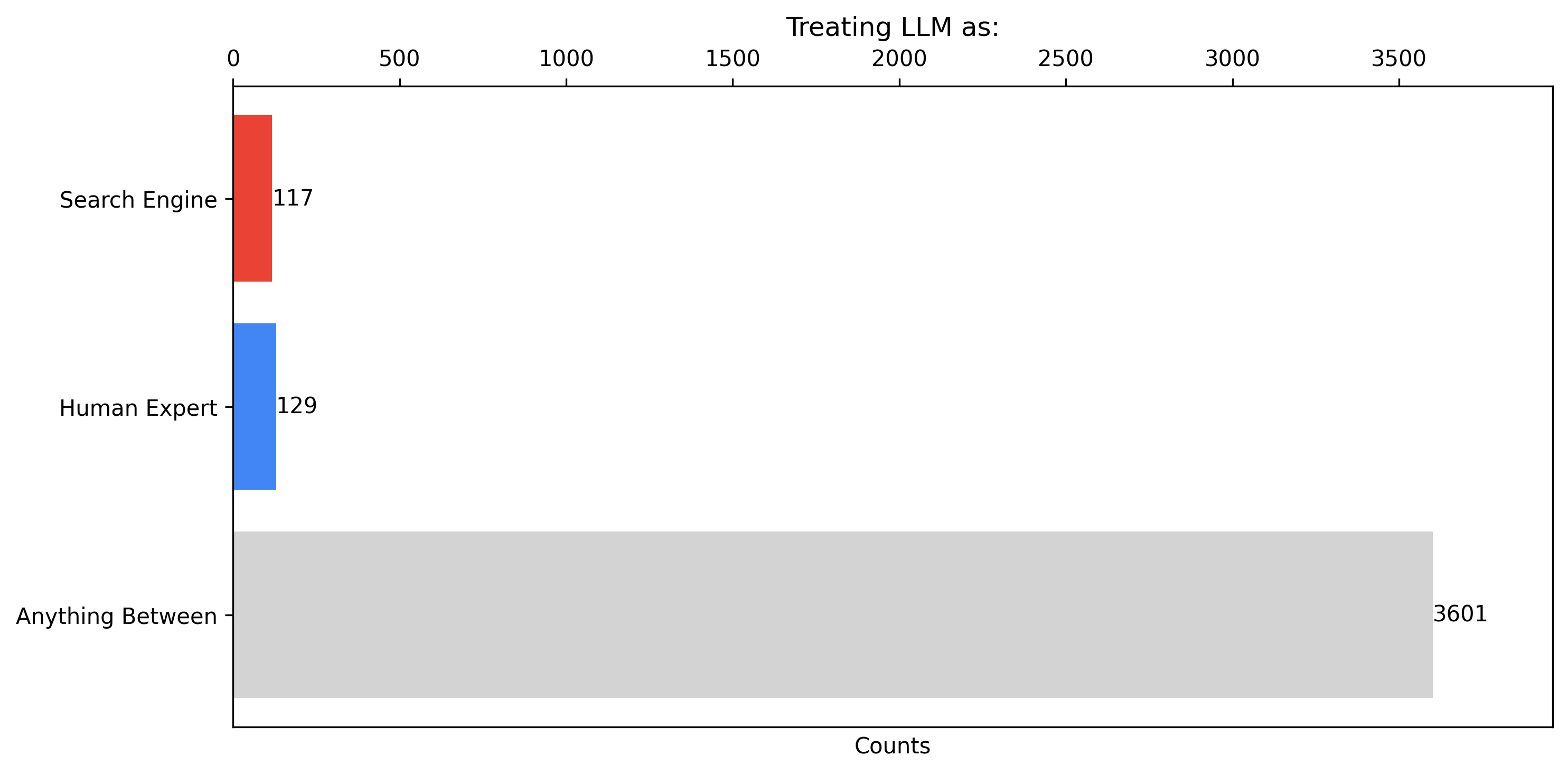}
    \caption{Proportion of Queries Seeking Personalized Actionable Advice by Treating the Model as a Human Expert (blue) and General Legal Information by Treating the Model as a Search Engine (red)}
    \label{fig:query-distribution}
\end{figure}

\section{Limitations and Discussion}
\label{discussion}

Before we engage in the discussion, it is necessary to outline the limitations of our analysis.

First, the experiment was primarily designed to explore and evaluate the potential applications of the GPT-4 model in the context of legal aid. The experiment was an internal effort of Frank Bold to bolster their non-profit activities and scale their legal clinic efforts. As such, no testable hypotheses were developed before the experiment. There are many variables which were not controlled. Arguably, our paper offers a great insight into actual and genuine legal issues people face and how they intuitively queried the GPT-4. At the same time, it means there is no demographic or other background available about the individual users, which could shed more light on the motivations and expectations of individual users.

Secondly, we developed the codes based on the existing literature and on iterative analysis of 200 randomly selected queries. The approach is reasonable within the confinement of our preliminary analysis. However, it gave us only limited validity of the results. Arguably, a more rigorous approach will lead to more nuanced categories, which will provide further (more detailed) insight into users' intuitive use of LLMs for legal aid.

Thirdly, we opted for zero-shot classification based on our codes and related definitions, which we used as prompts for GPT-4o. Zero-shot classification, while, as \citet{Savelka2023Effectiveness} put it, arguably 'unreasonably effective', has its limitations. We did not engage in any validation of the zero-shot classification provided by GPT-4o. The trends in user queries are clear. However, the precise numbers and ratios may be significantly different should the classification be done in other settings (a few-shot, manual, etc.)

Despite these limitations, our paper offers interesting—and, to our knowledge, previously unavailable—insights into intuitive user interaction with LLMs within legal aid. Our results provide several key takeaways.

The often-claimed risk of users disclosing their personal or other potentially sensitive information to LLMs or LLM-based applications is real and severe. When preparing the dataset for sharing between the researchers through the removal of personally identifiable information, the number of details disclosed by participants encountered by the first author (M.K.) was staggering. Especially within longer queries, users did not hesitate to include anything they deemed relevant as a factual context for their query. Within the nearly 30\% of queries containing facts, some users went to great lengths to describe anything that might have been relevant to answer their query. Such a level of detail would probably not have been necessary if human lawyers had provided the legal aid. We assume the uncertainty about the extent of relevant data and fear of failing to provide an essential piece of information is behind the oversharing. While the oversharing was exhibited only by the few users who included facts in their queries, it presents a clear risk and a troubling trend.

Furthermore, facts can provide context by hinting at, e.g., the necessity of using a specific statute. On the other hand, it shows that users aim to obtain answers personalized to their particular context. In principle, providing detailed facts can lead to models presenting more detailed answers but getting caught in the web of irrelevant or misrepresented information.

The majority of queries (nearly 65\%) sought information about the law, and the language of the query made it clear that they were not seeking legal advice or opinion. These queries focused on investigating legal concepts which users either encountered under their proper legal designations (e.g., easement) or were already familiar with through prior exposure (e.g., divorce proceedings). Seeking information about these concepts, including requests for the designation of related acts or identification of administrative or judicial bodies, was frequent. On the other hand, more than 35\% of queries sought what would constitute legal advice or opinion. Queries seeking LLM's advice on selection from possible procedural options, estimates on related costs, or chances of success were frequent. These answers would require detailed and first-hand knowledge about the inner workings of the courts in specific jurisdictions. As such, the user queries often manifested unfeasible expectations.

The overwhelming trend is that users did not maintain control over the LLM's answer to their query. Queries were mainly -- in more than 71\% cases -- open-ended and exploratory. That means that users imposed no further restrictions on the model's answer, e.g. through requirements for specific features or structural parts appearing in the answer or through queries requiring the step-by-step answer. Maintaining control limits the action space available to the model when formulating a response to a query. Once users grant control to LLM, it becomes more prone to hallucinations and providing irrelevant information. Skimming through these answers to open-ended questions may instil in users a false sense of competence. The answer may appear relevant. However, the failure to maintain control over the answer leaves users ultimately vulnerable. The trend is especially troubling in the context of legal aid and legal self-help.

Our findings are mostly in line with the findings of \citet{Hagan2024} and \citet{Cheong2024} and provide additional and more robust context. \citet{Hagan2024} hypothesized that people would use AI tools to deal with life and legal problems with increasing frequency and prominence. The experiment we outlined points towards the same conclusion. Additionally, Hagan hypothesized that many people will over-rely on AI tools to explain the law, even when encountering many disclaimers, warnings and real-life evidence of the contrary. Our analysis cannot offer a clear conclusion on over-reliance. Still, the findings suggest that users tend to intuitively engage LLMs with queries hinting at their ability to replace legal advice provided by human experts. \citet{Cheong2024} synthesized several principles which LLMs should follow to avoid providing a legal opinion. Our analysis suggests that these concerns are valid, and such principles may be necessary. The queries often suggest that advice or opinion was the primary need behind the query, but it was also a direct expectation.

The coding scheme we developed allows us to describe two end-of-spectrum approaches to users communicating with LLMs to seek legal aid. They can provide the model with facts of their case, use language clearly stating that they desire advice and opinion, not just information, and use open-ended questions, giving control over the answer to the model. Users whose queries combine these three factors treat LLM as if it were a human expert. On the other end of the spectrum are queries which provide no facts, use language clearly stating the user is interested in obtaining information about the law, and impose further limitations on the structure of the answer. Users whose queries combine these three factors treat LLM as if it were a sophisticated search engine. 

Interestingly enough, these extreme ends of the spectrum remain almost unpopulated. We identified only 117 queries (3.04\%) in which the users treated the model as if it were a sophisticated search engine. On the other end of the spectrum, we identified 129 queries (3.35\%) in which the users treated the model as if it were a human expert. Anything in between, where different approaches to querying, considerations, risks, and benefits intertwine, is poorly understood.

\section{Conclusion and Future Work}
\label{future}
While LLMs pose risks when misused, especially in high-stake contexts, they are here to stay. It does not seem reasonable to assume that laypeople will not use these tools for legal self-help just because they are informed they should avoid using them. Personalized legal services are prohibitively expensive even to the middle class, not to mention low-income and marginalized people. Warnings may lead users to be more careful when dealing with LLMs but will not prevent them from querying various models when seeking information. LLMs are here to stay and will be used - that much is clear from anecdotal evidence and the experiment introduced in the paper. 

Two strategies can be leveraged to address these issues - one on the side of users and one on the side of providers. There must be a significant increase in AI literacy \citet{Ng2021} throughout the society. As basic computer literacy became part of citizens' regular background knowledge, AI literacy must also become the norm. Future work should investigate laypeople's and professionals' intuitive use of LLMs, provide further insight, and offer strategies to increase their literacy to manage expectations and encourage responsible behaviour. On the other hand, users cannot be the only ones bearing the grunt of the widespread use of AI. Specific safeguards should be developed and implemented to mitigate the risks reasonably. These two efforts must go hand in hand.

Additionally, the experiment employed retrieval-augmented generation. FB's internal evaluation of the results suggests that users expressed varying satisfaction levels when receiving answers with different augmentation levels or using different augmentation methods. More robust experiments in this direction should follow. A layered approach forcing LLMs to consider specific—either expertly prepared or selected—documents and contexts could be an answer to some of the issues related to hallucinations and factuality.

The paper analysed user queries submitted by users Frank Bold, the Czech expert group providing for-profit and non-profit legal services. The experiment led 1,252 users to submit 3,847 queries to GPT-4. We classified queries based on whether or not they contained facts of the case, whether users required legal information or advice, and whether users posited open-ended questions or imposed further requirements on the structure of the model's answer. We used zero-shot classification by GPT-4o to classify the user queries. We have shown that users mosly do not provide LLMs with facts (70.05\% over 29.95\%), request information (as opposed to advice; 64.93\% over 35.07\%), and ask open-ended questions, granting control over the reply to the model (71.43\% over 28.57\%). We offered a unique insight into the intuitive use of LLMs. However, more detailed analysis and more rigorous experiments and surveys are required.

\section*{Acknowledgment}
Frank Bold's experiment and the work of M.K. were supported by the Ministry of Justice of the Czech Republic. J.H. was supported by the Technology Agency of the Czech Republic within the project CZDEMOS4AI (TQ12000040).

\bibliography{TEXTbibliography}

\end{document}